\documentclass{ws-procs975x65}

\usepackage[T1]{fontenc}
\usepackage{ae,aecompl}

\usepackage{amsmath,amsfonts,amssymb,mathrsfs}

\usepackage{tikz}
\usetikzlibrary{calc}
\usepackage{pgfplots}

\usepackage{xspace}

\usepackage{hyperref}
\hypersetup{colorlinks=true,linkcolor=blue,pdfstartview=XYZ}
\hypersetup{
colorlinks=false,
citecolor=green,
linkcolor=blue,
urlcolor=purple
}

\usepackage{xcolor}

\usepackage[titletoc]{appendix}
\usepackage{cleveref}

\usepackage{subfigure}

\usepackage{array}

\usepackage{moresize}

\def\beq{\begin{equation}}
\def\eeq{\end{equation}}
\def\bea{\begin{eqnarray}}
\def\eea{\end{eqnarray}}

\def \ti {\widetilde}

\newcommand{\Ups}{\Upsilon}

\newcommand{\rref}[1]{(\ref{#1})}
\newcommand{\di}{\mathrm{d}}

\newcommand{\lla}{\left\langle}
\newcommand{\rra}{\right\rangle}

\newcommand{\Scal}{\mathcal S}
\newcommand{\Acal}{\mathcal A}

\newcommand{\Ocal}{\mathcal O}

\newcommand{\Hcal}{\mathcal H}

\newcommand{\Pcal}{\mathcal P}

\newcommand{\tdev}{\mbox{\Large $\cdot$}}

\newcommand{\Mq}{{\cal M}_4}

\def \ti {\widetilde}

\begin{document}

\title{The Geodesic Light-Cone Coordinates, \\ an Adapted System for Light-Signal-Based Cosmology}

\vspace{-0.2cm}

\author{Fabien Nugier}

\address{Glenco Group, Dipartimento di Fisica e Astronomia, Universit\`{a} di Bologna, \\ viale B. Pichat 6/2 , 40127, Bologna, Italy
\\
E-mail: fabien.nugier@gmail.com\\
}

\vspace{-0.2cm}

\begin{abstract}
Most of cosmological observables are light-propagated. I will present coordinates adapted to the propagation of null-like signals as observed by a geodesic observer. These ``geodesic light-cone (GLC) coordinates'' are general, adapted to calculations in inhomogeneous geometries, and their properties make them useful for a large spectrum of applications, from the estimation of the distance-redshift relation, the average on our past light cone, the effect of the large-scale structure on the Hubble diagram, to weak lensing calculations.

\begin{center}
This document is a proceeding prepared for the \textsc{Fourteenth Marcel Grossmann Meeting}. \\
\end{center}

\end{abstract}

\vspace{-0.2cm}


\bodymatter


\setcounter{equation}{0}

\small

\section{Motivations}
\label{sec1}

The more observational precision increases, the more inhomogeneous our Universe looks. Supernov\ae\ \cite{Conley:2011ku}, lensing \cite{schneider2006gravitational,Amendola:2012ys,2009arXiv0912.0201L}, and other light-propagated observables will soon encounter associated complications
\cite{Clarkson:2011br,Fleury:2013sna,Lavinto:2015iba,AppEt1933a,Buchert:2011sx,Buchert:2011yu,
Clarkson:2011zq,Kolb:2011zz,Coley:2005ei}. 
We present coordinates first developed to simplify averages of scalars on our past light-cone \cite{P1}, next used to estimate the effect of inhomogeneities on luminosity distance \cite{P2,P4,Marozzi:2014kua,ErratumMarozzi2014}, Hubble diagram \cite{P3,P5}, and recently applied to lensing quantities \cite{P6} (and illustrated in a Lema\^itre-Tolman-Bondi model), following this historical order.

\vspace{-0.1cm}

\section{The geodesic light-cone coordinates}
\label{sec2}

We define a light-cone adapted metric (close to ``\emph{observational coordinates}''\cite{Maartens1,1985PhR...124..315E}, but different \cite{NugierThesis}) composed of 6 arbitrary functions ($\Ups$, $U^a$, $\gamma_{ab}$) and totally gauged fixed\,:
\beq
\di s_{\rm GLC}^2 = \Ups^2 \di w^2 - 2\Ups \di w \di \tau + \gamma_{ab} (\di \ti{\theta}^a - U^a \di w) (\di \ti{\theta}^b - U^b \di w) ~~.
\eeq

\vspace{-0.8cm}

\begin{figure}[ht!]
\centering
\includegraphics[width=5cm]{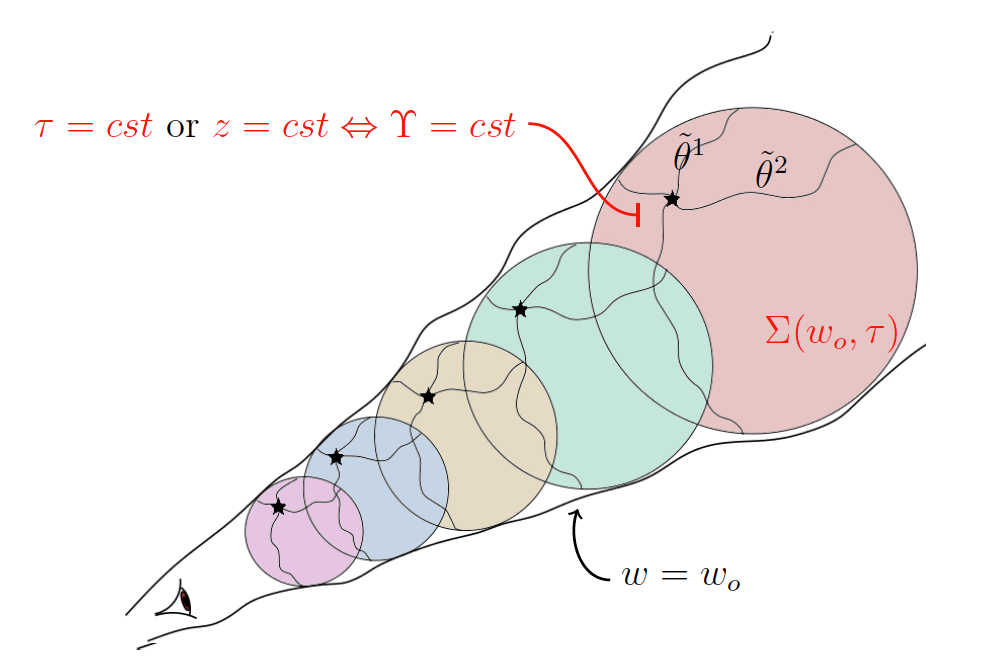} ~~
\includegraphics[width=5cm]{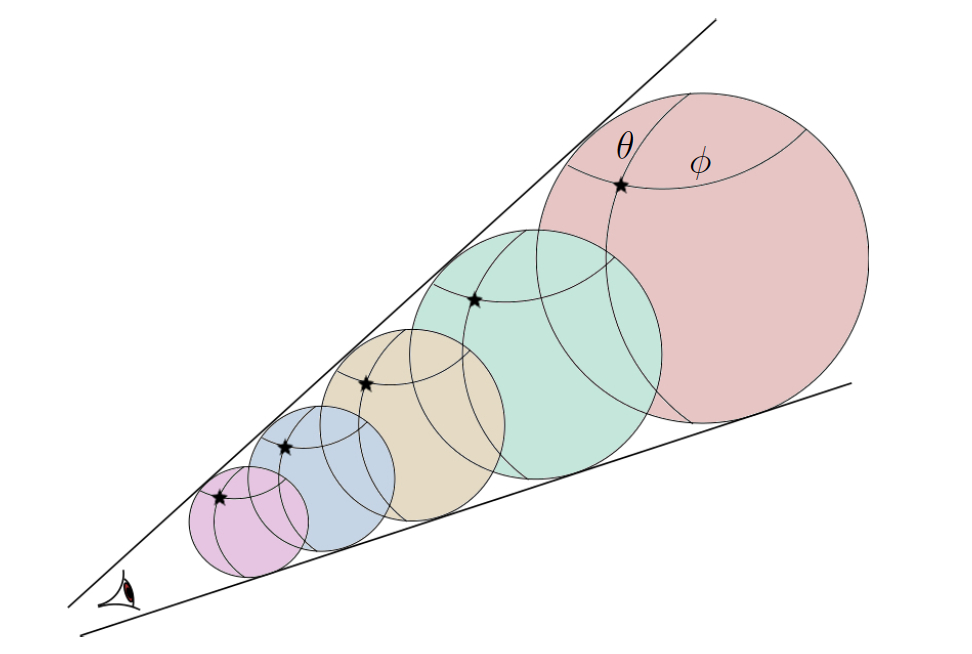}
\caption{\label{Fig1} \emph{Left\,:} Inhomogeneous light-cone parametrized by GLC coordinates. \emph{Right\,:} Homogeneous light-cone or an adapted system (like GLC coordinates) in an inhomogeneous geometry.}
\end{figure}

\vspace{-0.4cm}

This metric uses a null coordinate $w$ defining past light cones, the proper time of a geodesic observer $\tau$, and angles $\tilde{\theta}^a$ that photons keep along their path orthogonal to a 2-spheres $\Sigma(w,\tau)$ of constant time in our past light-cone (see Fig. \ref{Fig1}). In the FLRW limit\,: $w = \eta + r$ (conformal time, radius), $\tau=t$ (cosmic time), $(\tilde{\theta}^1,\tilde{\theta}^2) = (\theta,\phi)$, $\Ups = a(t)$, $U^a = 0$, $\gamma_{ab} = a^2 r^2 (1,\sin^2\theta)$. In general \cite{NugierThesis} $\Ups$ is like an inhomogeneous scale factor (lapse function), $U^a$ like a shift-vector and $\gamma_{ab}$ is the metric inside $\Sigma(w,\tau)$. We can notice two direct simplifications in GLC which, combined together, give the distance-redshift relation\,:
\vspace{-0.3cm}
\bea
\label{redshift}
& \mbox{Redshift\,:} ~~~ (1+z_s) = \Ups(w_o, \tau_o, \ti \theta^a) / \Ups(w_o, \tau_s, \ti \theta^a) ~~, \\
\label{AngDist}
& \mbox{Angular distance\,:} ~~~ d_A = \gamma^{1/4} (\sin\tilde{\theta}^1)^{-1/2} ~~\mbox{with}~ \gamma \equiv \det(\gamma_{ab}) = \frac{|\det (g_{\rm GLC})|}{\Ups^2} ~.~~
\eea
These coordinates share similarities with historical ones such as ``observational coordinates" \cite{saunders_observations_1968,saunders_observations_1969,Maartens1,1985PhR...124..315E}, (see elements of comparison in Ref.\cite{NugierThesis}) or the ``optical coordinates" \cite{1938RSPSA.168..122T}.

\section{Simplification of light-cone averages}
\label{sec3}

The light-cone average\cite{P1} of a scalar $S$ (e.g. $d_L ~,~ d_L^{-2}$) is in general given by $\lla S \rra (V_0,A_0) = I(S;V_0,A_0) / I(1;V_0,A_0)$ and we define the average integral to be gauge invariant, invariant under $(A,V) \rightarrow (\tilde{A}(A),\tilde{V}(V))$ and general coordinate transformations\,:
\beq
I(S;V_0,A_0) = \int_{\Mq} \di^4x \sqrt{-g} ~ D(V_0-V) D(A-A_0) ~ {\mathcal N}(V,A,\partial_\mu) ~ S(x) ~~,
\eeq
where ${\mathcal N}(V,A,\partial_\mu)$ is a normalization, $D(X) = \delta_D(X)$ or $\Theta(X)$ (Heaviside function)\,:

\vspace{-0.5cm}

\begin{table}[ht!]
\begin{center}
\begin{tabular}{| c | m{3cm} | m{3cm} | m{3cm} |}
\hline
$\mbox{Average}_{~\delta_D}^{~\Theta}$ & \centering $\lla S \rra_{V_0}^{A_0}$ & \centering $\lla S \rra_{A_0}^{V_0}$ & ~~~~~ ~~ $\lla S \rra_{V_0,A_0}$ \\
Illustration & ~~ \includegraphics[width=2.3cm]{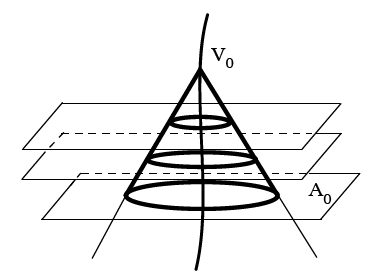} & ~~ \includegraphics[width=2.3cm]{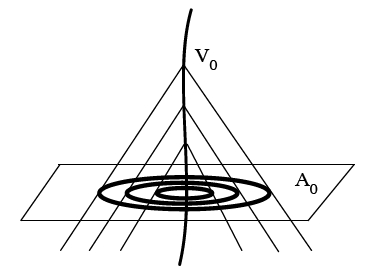} & ~~ \includegraphics[width=2.3cm]{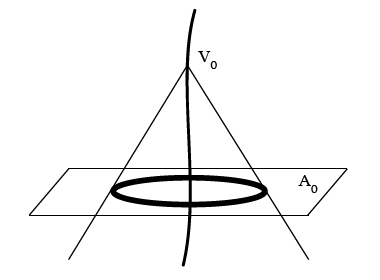} \\
${\mathcal N}(V,A,\partial_\mu)$ & \centering $\frac{|\partial_\mu V \partial^\mu A|}{\sqrt{-\partial_\nu A \partial^\nu A}}$ & \centering $\sqrt{-\partial_\mu A \partial^\mu A}$ & ~~~~~ ~ $|\partial_\mu V \partial^\mu A|$ \\
\hline
\end{tabular}
\end{center}
\end{table}

\vspace{-0.5cm}

Among these 3 types of averages, $\lla S \rra_{V_0,A_0}$ is closer to physical observables as it averages over the deformed 2-sphere embedded in the light-cone $V=V_0$ and a spatial hypersurface $A = A_0$. In GLC coordinates (where $V \rightarrow w$, $A \rightarrow \tau$) we can simplify the average and use Eq. \rref{redshift} to get (with $\tau_z \equiv \tau (z_s, w_o, \ti \theta^a))$\,:
\beq
\langle S \rangle_{w_o, z_s} =  \left( \int \di^2 \ti \theta \sqrt{\gamma (w_o, \tau_z, \ti \theta^b)} \, S(w_o, \tau_z, \ti \theta^b) \right) \Big / \left( \int \di^2 \ti \theta \sqrt{\gamma (w_o, \tau_z, \ti \theta^b)} \right) ~~,
\eeq
allowing us to average scalars on the sky, at a certain redshift.

\section{Distance-redshift relation at $\Ocal(2)$}
\label{sec4}

The GLC metric enables the computation of $d_L(z)$ at $\Ocal(2)$ in the Newtonian gauge (NG)\,:
\beq
\di s_{\rm NG}^2 = a^2(\eta) \left[ -(1+ 2 \Phi) \di \eta^2  + (1- 2 \Psi) ( \di r^2 + r^2 [\di \theta^2 + \sin^2 \theta \, \di \phi^2]) \right]
\eeq
with $\Phi = \psi + \frac12 \phi^{(2)}$ , $\Psi = \psi + \frac12 \psi^{(2)}$ the gauge invariant Bardeen potentials (no matter shear at $\Ocal(1)$, or see Ref.\cite{Marozzi:2014kua}). Establishing the full transformation between GLC and NG coordinates at $\Ocal(2)$ in (scalar) perturbations, we get $(\tau, w,\ti{\theta}^1,\ti{\theta}^2) = {\rm func}(\eta, r, \theta, \phi)$ and $(\Ups, U^a, \gamma^{ab}) = {\rm func}(\psi,\psi^{(2)},\phi^{(2)})$ which allows us to compute $d_L$ to $\Ocal(2)$ (using Eq. \rref{AngDist})\,:
\beq
d_L(z_s, \theta^a) = d_{L}^{FLRW}(z_s) \left( 1 + \delta_S^{(1)}(z_s, \theta^a) + \delta_S^{(2)}(z_s, \theta^a) \right) ~~.
\eeq

The first order in scalar perturbations is given by\,:
\beq
\delta_S^{(1)}(z_s, \theta^a) = \left( 1 - \frac{1}{\Hcal_s \Delta \eta} \right) J - \frac{Q_s}{\Delta \eta} - \psi_s^{(1)} - {\rm Lensing}^{(1)} ~,
\eeq
with $\Hcal_s = a'(\eta_s)/a(\eta_s)$, $\Delta \eta = \eta_o - \eta_s^{(0)}$, containing (Integrated) Sachs-Wolfe ([I]SW), Doppler, and lensing (convergence) effects \,:
\bea
& J  = ([\partial_+ Q]_s - [\partial_+ Q]_o) - ([\partial_r P]_s - [\partial_r P]_o) \sim {\rm SW} + {\rm ISW} + {\rm Doppler} ~~, \\
\medskip
& {\rm Lensing}^{(1)} = \frac12 \nabla_a \tilde{\theta}^{a (1)} = \int_{\eta_s^{(0)}}^{\eta_o} \frac{\di \eta}{\Delta\eta} ~ \frac{\eta - \eta_s^{(0)}}{\eta_o - \eta} \Delta_2 \psi (\eta, \eta_o-\eta, \theta^a) ~~, \nonumber
\eea
where `$o$' (`$s$') denote quantities evaluated at the observer (source), and we defined\,:
\beq
Q(\eta_+, \eta_-, \theta^a) = \int_{\eta_+}^{\eta_-} \di x~ \hat{\psi}(\eta_+,x,\theta^a) ~~,~~ P(\eta, r, \theta^a) = \int_{\eta_{in}}^\eta \di \eta'~ \frac{a(\eta')}{a(\eta)} \psi(\eta',r,\theta^a) ~~.
\eeq

Similarly obtained $\Ocal(2)$ corrections contain (see Refs.\cite{P4,NugierThesis})\,:
\begin{itemize}
  \setlength{\itemsep}{-5pt}
  \setlength{\parskip}{0pt}
  \setlength{\parsep}{0pt}
\item Two dominant terms\,: $(\mbox{Doppler})^2$, $(\mbox{Lensing})^2$, \\
\item Combinations of $\Ocal(1)$-terms\,: $\psi_s^2$, $(\mbox{[I]SW})^2$, $\mbox{[I]SW} \times \mbox{Doppler}$, $(\psi_s, Q_s) \times $ (Lensing, [I]SW, Doppler), \\
\item Genuine $\Ocal(2)$-terms\,: $\psi_s^{(2)}$, $Q_s^{(2)}$, $\mbox{Lensing}^{(2)} \! = \! \frac12 \nabla_a \tilde{\theta}^{a (2)}$, \\
\item New integrated effects\,: $\frac{1}{4 \Delta\eta} \int_{\eta_s^{(0)+}}^{\eta_s^{(0)-}} \di x~ \left[4 \hat{\psi} ~ \partial_+ Q + \hat{\gamma}_{0}^{ab} ~ \partial_a Q ~ \partial_b Q \right] (\eta_s^{(0)+},x,\theta^a)$, \\
\item Angle deformations\,: $(\gamma_0)_{ab} \partial_+ \tilde{\theta}^{a (1)}\partial_- \tilde{\theta}^{b (1)} ~~,~~ \partial_a \tilde{\theta}^{b (1)}\partial_b \tilde{\theta}^{a (1)}$ ~, \\
\item Redshift perturbations from Eq. \rref{redshift}, involving transverse peculiar velocities\,: \\ $\gamma_0^{ab} \left( \partial_a P ~ \partial_b P ~,~ \partial_a Q ~ \partial_b Q ~,~ \partial_a Q ~ \partial_b P \right)$ , 
$\partial_+ \int_{\eta_+}^{\eta_-} \di x  \left[ 4 \hat{\psi} ~ \partial_+ Q + \hat{\gamma}_0^{ab} \partial_a Q ~ \partial_b Q \right]$ , 
$\int_{\eta_{in}}^\eta \di \eta'~ \frac{a(\eta')}{a(\eta)} \partial_r \left[\left( \partial_r P \right)^2 + \gamma_0^{ab} \partial_a P ~ \partial_b P \right]$ , \\
\item Other important terms\,: Lens-Lens coupling, corrections to Born approximation.
\end{itemize}

These $\Ocal(2)$ results were confirmed recently\cite{Fanizza:2015swa} working directly in terms of GLC coordinates rather than NG. An independent derivation \cite{2014CQGra..31t2001U1,2014CQGra..31t5001U2} lead to similar results, but a rigorous comparison with Ref.\cite{P4} is still lacking. Results were also given for vector/tensor perturbations\cite{P4} (Poisson gauge) and extended to the case with $\Ocal(1)$ anisotropic stress \cite{Marozzi:2014kua,ErratumMarozzi2014}.

\section{Effects of large-scale structure on the Hubble diagram}
\label{sec5}

In Sec. \ref{sec4} we expressed $d_L$ in terms of $(\psi,\psi^{(2)},\phi^{(2)})$, hence it needs a description of the Bardeen potentials at $\Ocal(1,2)$. We decompose the first order gravitational potential $\psi$ in Fourier modes and denote by $\overline{(...)}$ the ensemble (or stochastic) average over perturbations. The $\Ocal(2)$ potentials can be related to $\psi$ by\cite{Bartolo:2005kv}\,: $\psi^{(2)}, \phi^{(2)} \propto \nabla^{-2} (\partial_i \psi \partial^i \psi) ~,~ \partial_i \psi \partial^i \psi$. Hence we can combine light-cone and stochastic averages in order to study the effect of statistical perturbations on the whole sky. For example, the (trivial) average of $\psi^2$ gives\,: $\overline{\lla \psi_s^2 \rra} = \int_{0}^{\infty} \frac{\di k}{k} \Pcal_{\psi}(k)$ where $\Pcal_\psi (k) \equiv (k^3 / 2 \pi^2) |\psi_k(\eta)|^2 = (3 / 5)^2 A (k/k_0)^{n_s-1} T^2(k) \, g^2(z)$ is the power spectrum describing perturbations. At linear order, with $A$, $n_s$, $k_0$ taken from WMAP, $T(k)$ is a transfer function\cite{Eisenstein:1997ik} including a baryonic component (Silk damping), and $g(z)$ is the growth factor describing the recent time evolution of perturbations.
In CDM we get exactly the spectral coefficients coming from each correction of $d_L(z_s, \theta^a)$ described in Sec. \ref{sec4}\,:
$\overline{\lla d_L \rra} = \int_0^\infty \frac{\di k}{k} \Pcal_{\psi}(k) \, C(k \Delta\eta)$. We do the same in $\Lambda$CDM, with reasonable assumptions to simplify integrations \cite{P5}, and also with a non-linear power spectrum \cite{Smith:2002dz,Takahashi:2012em}.

Like $d_L$ one can also average the flux $\Phi = L / (4 \pi d_L^2) \simeq \Phi_0 + \Phi_1 + \Phi_2$. We get\,:
\begin{itemize}
  \setlength{\itemsep}{2pt}
  \setlength{\parskip}{0pt}
  \setlength{\parsep}{0pt}
\item $\overline{\langle d_L^{-2} \rangle} \equiv (d_L^{FLRW})^{-2} \left[1 + f_\Phi(z) \right]$ where $f_\Phi(z) \simeq f(z) \int_0^{\infty} \frac{\di k}{k} \left(\frac{k}{  {\cal H}_0}\right)^2 \Pcal_{\psi}(k)$ ~~, \\
\item $\overline{\langle d_L \rangle}(z) = d_L^{FLRW} \left[1 + f_d(z) \right]$ with $f_d = -(1/2) f_{\Phi} + (3/8) \overline{\langle \left(\Phi_1/\Phi_0\right)^2 \rangle}$~~.
\end{itemize}

\medskip

\begin{figure}[ht!]
\centering
\includegraphics[width=5cm]{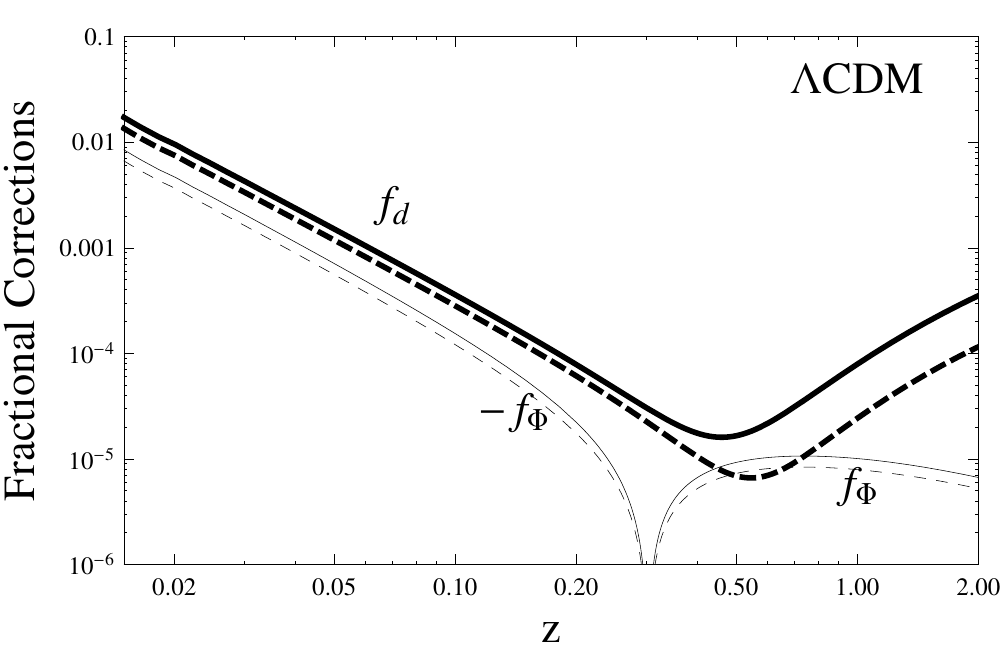} ~~~~
\includegraphics[width=5.1cm]{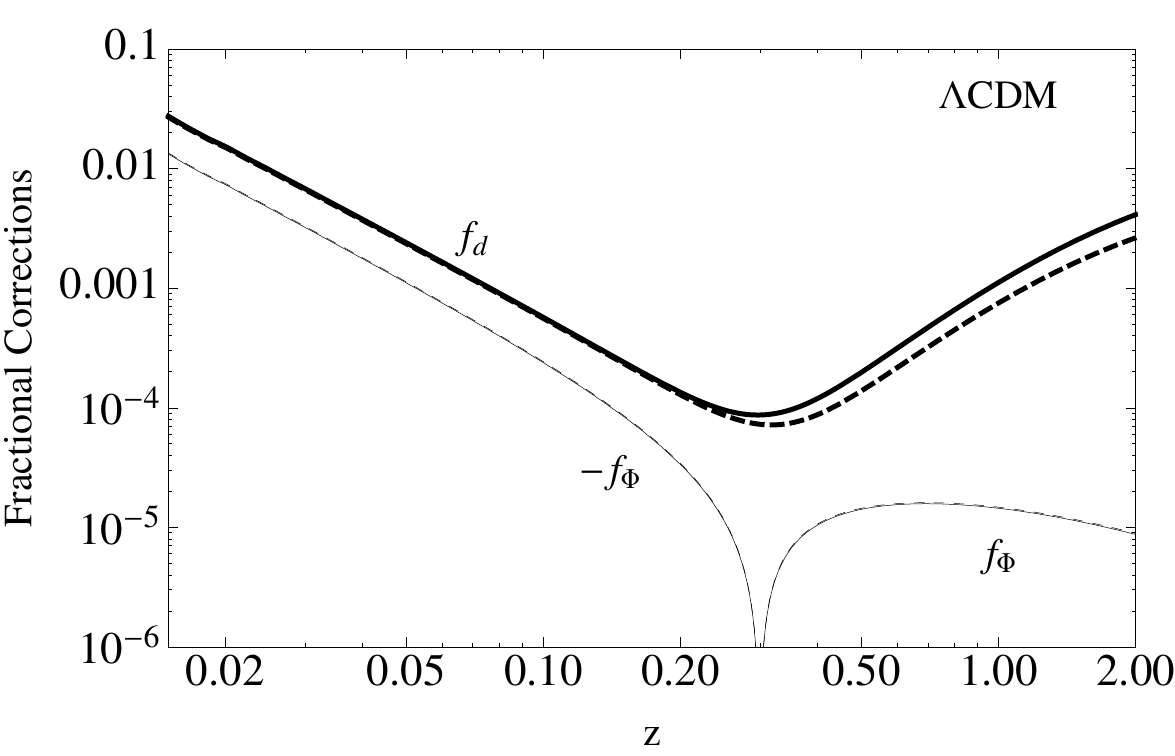}
\caption{\label{Fig2} Corrections $f_\Phi(z)$ and $f_d(z)$ for a linear (\emph{Left}) and non-linear (\emph{Right}) power spectrum.}
\centering
\end{figure}

Corrections to $d_L$ involve a flux variance dominated by peculiar velocity and lensing\,:
\scriptsize
\beq
\overline{\lla \left(\phi_1 / \phi_0 \right)^2 \rra}/4 = \overline{\lla (\delta_S^{(1)})^2 \rra} \simeq \left( 1 - \frac{1}{\Hcal_s \Delta\eta} \right)^2 \left\{ \left[\overline{\langle  ([\partial_r P]_s)^2 \rangle} + \overline{\langle  ([\partial_r P]_o)^2  \rangle}\right] + \overline{\left\langle \left({\rm Lensing}^{(1)}\right)^2 \right\rangle} \right\} ~~, \nonumber
\eeq
\small
as shown in Fig. \ref{Fig2} for realistic (non-)linear power spectra \cite{P5}. It turns out that the luminosity flux is minimally affected by lensing w.r.t. other scalar observables at large redshift. This calculation can be seen as a check at $\Ocal(2)$ of Weinberg's argument of flux conservation \cite{1976ApJ...208L...1W} and has also been confirmed by recent papers through different approaches \cite{Kaiser:2015iia,2005ApJ...632..718K,2015JCAP...07..040B}.

\begin{figure}[ht!]
\centering
\includegraphics[width=5cm]{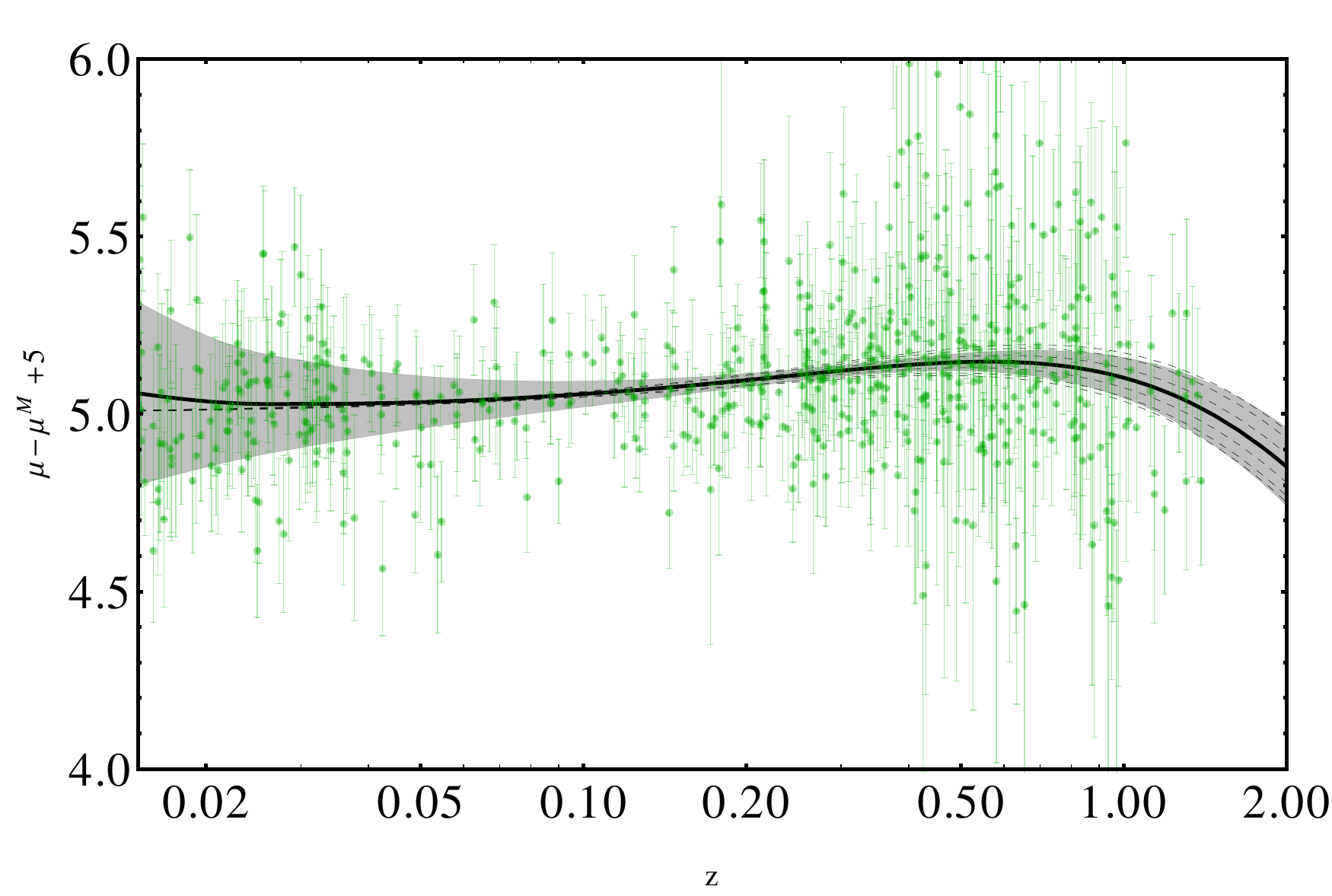} ~~
\includegraphics[width=5.3cm]{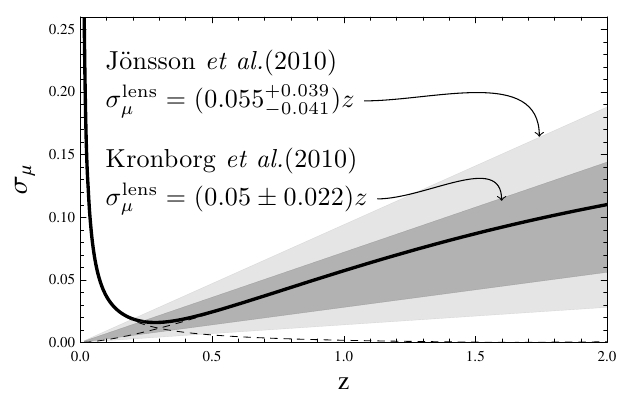}
\caption{\label{Fig3} \emph{Left\,:} Theoretical dispersion from power spectrum vs Union 2 data. \emph{Right\,:} Dispersion $\sigma_{\mu}$ from peculiar velocities and lensing (solid line) compared to experimental estimates (grey areas).}
\centering
\end{figure}

Similarly, we can get the average/dispersion of the distance modulus\,:
\beq
\overline{\langle \mu \rangle} = \mu^{\rm FLRW} - 1.25 (\log_{10}e) \Big[ 2f_\Phi - \overline{\langle \left(\Phi_1/\Phi_0\right)^2 \rangle} \Big] ~~,~~ \sigma_\mu = 2.5 (\log_{10} e) \sqrt{\overline{\langle \left(\Phi_1/\Phi_0\right)^2 \rangle}} ~~.
\eeq
Compared to the Union 2 data and using a non-linear power spectrum in $\Lambda$CDM (Fig. \ref{Fig3}, Left), we find that peculiar velocities explain well the scatter at small $z$ and that lensing explains only part of the scatter at large $z$. Finally, we can compare our dispersion on the Hubble diagram with the experimental estimations coming from lensing\cite{Jonsson:2010wx,Kronborg:2010uj} (Fig. \ref{Fig3}, Right). We find that the total effect is well fitted by Doppler ($z \leq 0.2$) + lensing ($z > 0.3$) effects and that the lensing prediction is in great agreement with experiments so far.

\section{Jacobi map and weak lensing}
\label{sec6}

We now consider lensing\cite{P6}, motivated by Sec. \ref{sec5} and recent work\cite{DiDio:2014lka} on galaxy number counts in GLC. The relative separation of two neighbour light rays simultaneously emitted from a source $S$ and converging to an observer $O$ follows the geodesic deviation eq. (GDE)\,: $\nabla_\lambda^2 \xi^\mu = R^{\mu}_{~\,\alpha\beta\nu} k^\alpha k^\nu \xi^\beta$ with $\nabla_\lambda\equiv {\rm D}/\di \lambda \equiv k^\mu \nabla_{\mu}$, $k^\mu$ the photon momentum, $\lambda$ an affine parameter along the photon path, and $\xi^\mu$ an orthogonal displacement w.r.t. to the rays.
We project the GDE on the Sachs basis $\{ s^\mu_A \}_{A=1,2}$ (two zweibeins with flat index $A=1,2$)\,: $g_{\mu\nu} s^\mu_A s^\nu_B = \delta_{AB} ~,~ s^\mu_A u_\mu = 0 ~,~ s^\mu_A k_\mu = 0 ~,~ \Pi^\mu_\nu\nabla_\lambda s^\nu_A = 0$ ; with $u_\mu$ the peculiar velocity of comoving fluid ($S$, $O$ comoving too), $\Pi^\mu_\nu$ a ``screen" projector orthogonal to $u_\mu$ and $u_\mu+\left( u^\alpha k_\alpha \right)^{-1} k_{\mu}$.
We define the Jacobi map $J^A_B$, from the observed sky angle $\bar{\theta}_o^A$ to $\xi^A \equiv \xi^\mu s^A_\mu$, by $\xi^A(\lambda) = J^A_B(\lambda,\lambda_o) \, \bar{\theta}_o^A$. Projected quantities $\xi^A$ and $R^A_B \equiv R_{\alpha\beta\nu\mu}k^\alpha k^\nu s^\beta_B s^\mu_A$ (optical tidal matrix) bring us the Jacobi equation (see e.g. Refs.\cite{Fleury:2013sna,Fanizza:2013doa})\,:
\bea
\label{JacobiEq}
& \frac{\di^2}{\di \lambda^2} J^A_B (\lambda , \lambda_o) = R^A_C (\lambda) \, J^C_B (\lambda, \lambda_o) ~~, \\
\label{JacobiEqIC}
& \mbox{with I.C.\,:} ~~~ J^A_B(\lambda_o, \lambda_o) = 0 ~~~,~~~ \frac{\di}{\di \lambda} J^A_B (\lambda_o,\lambda_o) = (k^\mu u_\mu)_o \, \delta^A_B ~~.
\eea
A direct resolution of Eq. \rref{JacobiEq} gives the angular distance of the source $d_A(\lambda_s) \equiv \sqrt{d S_s / d^2 \Omega_o} = \sqrt{\det J^A_B(\lambda_s,\lambda_o)}$.
Also, the (unlensed) angular position of the source $\bar\theta^A_s$ and the observed lensed position $\bar\theta^A_o$ (of the image) are given by\,: $\bar\theta^A_s=\left(\xi^A / \bar d_A\right)_s$, $\bar\theta^A_o=\left( k^\mu\partial_\mu\xi^A / k^\mu u_\mu \right)_o$, where $\bar{d}_{A}$ is the homogeneous and isotropic background our model refers to.
This allows us to define the so-called amplification matrix as\,:
\beq
\Acal^A_B \equiv \frac{\di \bar{\theta}^A_s}{\di \bar{\theta}^B_o} = \frac{J^A_B (\lambda_s,\lambda_o)}{\bar d _A(\lambda_s)} = \left( \begin{array}{cc} 1 - \kappa - \hat{\gamma}_1 & - \hat{\gamma}_2 + \hat{\omega} \\ - \hat{\gamma}_2 - \hat{\omega} & 1 - \kappa + \hat{\gamma}_1 \end{array} \right) ~~,
\eeq
which defines the lensing quantities\,: $\kappa = 1-\text{tr}J^A_B / 2\bar d_A$ (convergence), $\hat\omega=|J^1_2-J^2_1| / 2\bar d_A$ (vorticity), $|\hat\gamma| \equiv \sqrt{(\hat\gamma_1)^2 + (\hat\gamma_2)^2} = \sqrt{(1-\kappa)^2 + \hat{\omega}^2 - \mu^{-1}}$ (shear) and $\mu \equiv 1/(\det\Acal)=\bar d_A^2 / \det J^A_B$ (magnification).

Let us now turn to the GLC coordinates and express these lensing quantities in it. First, the zweibeins are written as $s^\mu_A=(s^\tau_A,0,s^a_A)$ and $k^\mu \equiv \omega \Ups^{-1}\delta^\mu_\tau$ (with $\omega$ a pure constant). Second, the solution to Eqs. \rref{JacobiEq} and \rref{JacobiEqIC} is\,:
\beq
J^A_B(\lambda,\lambda_o) = s_a^A(\lambda)\,\left[ 2 u_{\tau} (\dot{\gamma}_{ab})^{-1} \right]_os^B_b(\lambda_o)
\eeq
where $(\ldots)^{\tdev} \equiv \partial_\tau (\ldots)$\,.
The angular distance and the magnification become\,:
\beq
d_A = 2 u_{\tau_o} (\gamma \gamma_o)^{1/4} \big / \sqrt{(\det \dot{\gamma}_{ab})_o} ~~~,~~~ \mu = \left(\bar d_A / d_A\right)^2 = \Phi / \bar{\Phi} ~~,
\eeq
involving $\bar{d}_A = a^2(\tau) r^2$ with $r = w - \int \di \tau / a(\tau)$ measured from the observer and $\Phi$ ($\bar{\Phi}$) the flux in the in(homogeneous) geometry. Expressions for the zweibeins can be obtained in the GLC coordinates\cite{P6}, but it is more convenient to compute the squared lensing quantities, combined with $s^A_as^A_b=\gamma_{ab}$ and $\epsilon_{AB}\,s^A_as^B_b=\sqrt{\gamma}\,\epsilon_{ab}$ ($\epsilon$ the anti-sym. symbol), to get\,:
\beq
\left( \begin{array}{c} \left( 1-\kappa \right)^2+\hat\omega^2 \\ \hat\gamma_1^2+\hat\gamma_2^2 \end{array} \right) = \left( \frac{u_{\tau_o}}{\bar d_A} \right)^2 \left( \left[ \frac{\gamma\,\dot\gamma_{ab}\gamma^{bc}\dot\gamma_{cd}}{\left(\det^{ab}\dot\gamma_{ab}\right)^2} \right]_o\gamma\,\gamma^{ad} \pm  \frac{2 \sqrt{\gamma\,\gamma_o}}{\left( \det^{ab}\dot\gamma_{ab} \right)_o} \right) ~~,
\eeq
We thus have general lensing quantities expressed with only 3 metric functions (of $\gamma_{ab}$), showing the great advantage of working in GLC coordinates.

The Jacobi Eq. \rref{JacobiEq} can be rewritten as a first order differential equation for the so-called deformation matrix\,:
\beq
\Scal^A_B\equiv \frac{\di J^A_C}{\di \lambda} (J^{-1})^C_B =\hat\theta\,\delta^A_B+\left(\begin{array}{cc}
\hat\sigma_1&\hat\sigma_2\\
\hat\sigma_2&-\hat\sigma_1
\end{array}\right) ~~,
\eeq
involving the optical scalars, $\hat\theta$ (expansion scalar) and $\hat\sigma \equiv \hat\sigma_1 + i \hat\sigma_2$ (shear scalar), and known as the Sachs equations\,:
\beq
\frac{\di \hat\theta}{\di \lambda}+|\hat\sigma|^2+\hat\theta^2 = \Phi_{00} ~~~~~,~~~~~ \frac{\di \hat\sigma}{\di \lambda}+2\hat\theta \hat\sigma = \Psi_0 ~~.
\eeq
The RHS terms are the Ricci and Weyl focusing and are defined as follows\,:
\beq
\Phi_{00}=-\frac{1}{2}R_{\alpha\beta} k^\alpha k^\beta ~~~,~~~ \Psi_0=\frac{1}{2}C_{\alpha\beta\mu\nu}k^\alpha k^\mu \Sigma^\beta \Sigma^\nu ~~,
\eeq
where $R_{\alpha\beta}$ is the Ricci tensor, $C_{\alpha\beta\mu\nu}$ the Weyl tensor, and $\Sigma^\mu \equiv s^\mu_1+is^\mu_2$.
As well as the amplification matrix, the deformation matrix simplifies in the GLC coordinates as\,:
\beq
\Scal^A_B = \frac{\di s^A_a}{\di \lambda} s^a_B = \frac{\omega}{2\Ups} s^a_A s^b_B\, \dot \gamma_{ab} ~~.
\eeq
Using $s^a_A s^b_A=\gamma^{ab}$ we get the optical scalars\,:
\beq
\hat\theta = \omega\,\frac{\gamma^{ab}\dot\gamma_{ab}}{4\Ups}=\frac{\omega}{4\Ups} \frac{\dot\gamma}{\gamma} ~~~~~,~~~~~ |\hat\sigma|^2 = \left(\frac{\omega}{4 \Ups} \frac{\dot\gamma}{\gamma}\right)^2 - \frac{\omega^2}{4\Ups^2}\,\frac{\det\dot\gamma_{ab}}{\gamma} ~~.
\eeq
We also get the Ricci and Weyl focusing in GLC coordinates\,:
\beq
\Phi_{00} = \frac{\omega^2}{4\Ups^2} \gamma^{ab} \, Y_{ab} ~~,~~ |\Psi_0| = \frac{\omega^2}{4\Upsilon^2} \sqrt{\left( \gamma^{ac}\gamma^{bd}+\gamma^{ad}\gamma^{bc}-\gamma^{ab}\gamma^{cd} \right) Y_{ab} Y_{cd}}~,
\eeq
where 
$Y_{ab} \equiv \ddot \gamma_{ab} - (\dot \Ups / \Ups) \dot\gamma_{ab} - (1/2) \dot\gamma_{ac}\gamma^{cd}\dot\gamma_{db}$ depends only on $\gamma_{ab}$, $\Ups$ and their time derivatives. This proves again the usefulness of GLC coordinates for lensing and it was illustrated\cite{P6} by the computation of lensing quantities in the case of an off-center observer in a Lema\^itre-Tolman-Bondi model (considering only the decaying mode).

\section{Conclusions}
\label{sec7}

We have shown that there are many advantages in using the GLC coordinates. They are indeed adapted to calculations involving light-propagation, they can also be used for weak lensing (where $\gamma_{ab}$ acts as a screen), and may help to get new predictions on cosmology or to study other aspects of lensing (e.g. lensing statistics).

\section*{Acknowledgments}
I want to thank the \textsc{Fourteenth Marcel Grossmann Meeting} for giving me the opportunity to share these properties of GLC coordinates in two parallel sessions and for giving me the occasion to write this document. My researches are supported by the project GLENCO lead by B. Metcalf, funded under the FP7, Ideas, Grant Agreement n. 259349.


\ssmall

\providecommand{\href}[2]{#2}\begingroup\raggedright\endgroup

\end{document}